\documentstyle[12pt]{article}
\begin{document}

\baselineskip=24pt plus 2pt
\hfill\hbox{NCKU-HEP/97-01}
\begin{center}

{\large \bf On the M{\o}ller's energy complex of
 the charged dilaton black hole }\\

\vspace{5mm}
I-Ching Yang  \footnote{E-mail:icyang@ibm65.phys.ncku.edu.tw},
Wei-Fui Lin  \footnote{E-mail:wfl@ibm65.phys.ncku.edu.tw},
and Rue-Ron Hsu  \footnote{E-mail:rrhsu@mail.ncku.edu.tw}
\vspace{5mm}

Department of Physics, National Cheng Kung University \\
Tainan, Taiwan 701, Republic of China \\

\end{center}
\vspace{5mm}

\begin{center}

{\bf ABSTRACT}
\end{center}
    Using M{\o}ller's energy complex , we obtain the 
energy distributions of GHS solution and dyonic dilaton black
hole solution in the dilaton gravity theory. 
It is confirmed that the M{\o}ller's energy complex
is indeed a 3-scalar under purely spatial transformation in these
energy distributions. Some interested properties of the energy
distribution of dyonic black hole are disscussed.

\vspace{2mm} 
\noindent
{PASC:04.20.-q, 04.50,+h} \\
{\it keywords:}{M{\o}ller's energy complex, dilaton gravity theory, GHS solution, dyonic dilaton solution}
\newpage
 
    The well - known Einstein's energy complex is the foremost
definition of energy complex. This idea comes from that the continuity
equation $ \frac{\partial T^\mu_\nu}{\partial x^\mu} = 0 $ can 
give out a conserved quantity in the absence of a gravitation field. 
However, when the gravitational field is present, the conservation law
will be generalized to be 
\begin{equation}
  \nabla_\mu T^\mu_\nu = \frac{1}{\sqrt{-g}} \frac{\partial (\sqrt{-g} T^\mu_\nu )}{\partial x^\mu} - \frac{1}{2} \frac{\partial g_{\mu\sigma}}{\partial x^\nu} T^{\mu\sigma} = 0  .
\end{equation}
It does not generally induce any conservation quantity. In order to 
determine the conserved total four-momentum, a particular coordinate
system where all the first derivatives of the $ g_{\mu\nu} $ are vanish
at some particular points must be chosen. However, the Einstein's energy complex does not supply
a physically satisfactory description of the energy distribution or of
the energy contained in a limited part of space. Due to the the condition
that the first - order derivative of $ g_{\mu\nu} $ must be equal to 
zero, one must specify the calculation at the large spatial
distance $ r $ from the system in Cartestian coordinate. Thus,
it has no meaning to speak of a definite localization of
gravitation field in this case~\cite{1}.
Another energy complex which has the property of transforming as a
3 - scalar density with respect to the group of purely spatial
transformations was defined by M{\o}ller~\cite{2,3}. This property was 
shown that the energy distribution will be the same while calculating
in different kinds of spatial coordinate (but the same time coordinate).
In this paper, we calculate the energy distributions of two black hole
solutions in dilaton gravity theory to examine the property of M{\o}ller
energy complex. 

    In the dilaton gravity theory in which the gravity is coupled to the electromagnetic
and dilaton fields can be described by the four - dimensional 
effective string action~\cite{4}. The action can be expressed as 
\begin{equation}
  I = \int d^4 x \sqrt{-g}  \left[ -R + 2(\nabla \phi)^2 + e^{-2\phi}F^2 \right] .
\end{equation} 
Grafinkle, Horowitz and Strominger found a charged dilaton 
black hole solutions(GHS solutions)~\cite{4} in a peculiar coordinate form.
Their solutions are
\begin{eqnarray}
  ds^2 &=& (1-\frac{2M}{\stackrel{\sim}{r}})dt^2 - \frac{1}{(1-\frac{2M}{\stackrel{\sim}{r}})}d\stackrel{\sim}{r}^2 - (1-\frac{\alpha}{\stackrel{\sim}{r}})\stackrel{\sim}{r}^2(d\theta^2 + sin^2\theta d\varphi^2) ,   \\
  \alpha & = & \frac{Q_e^2}{M}e^{2\phi_0} ,   \\
  e^{-2\phi} & = & e^{-2\phi_0} (1 - \frac{Q_e^2}{M \stackrel{\sim}{r}}e^{2\phi_0}) ,  \\
  F_{01} & = & \frac{Q_e}{\stackrel{\sim}{r}^2}e^{2\phi} .   
\end{eqnarray}
The properties of the GHS solutions are characterized by the mass $ M $, electric charge
$ Q_e $ and asympotic value of the dilaton $ \phi_0 $.
On the other hand, by using a standard spherical coordinate form
\begin{equation}
  ds^2 = \Delta^2dt^2 - \frac{\sigma^2}{\Delta^2}dr^2 - r^2d\theta^2 -  r^2sin^2\theta d\varphi^2 ,
\end{equation}
Cheng, Lin and Hsu~\cite{5} had obtained the dyonic dilaton black hole solutions (CLH solutions)
\begin{eqnarray}
  \Delta^2 & = & 1 - \frac{2M}{r^2}\sqrt{r^2+\lambda^2} + \frac{\beta}{r^2} ,  \\
  \sigma^2 & = & \frac{r^2}{r^2+\lambda^2} ,  \\
  \lambda & = & \frac{1}{2M}(Q_e^2e^{2\phi_0} - Q_m^2e^{-2\phi_0}) ,  \\
  \beta & = & (Q_e^2 e^{2\phi_0} + Q_m^2 e^{-2\phi_0}) , \\
  e^{2\phi} & = & e^{2\phi_0}(1 - \frac{2\lambda}{\sqrt{r^2+\lambda^2} + \lambda}) ,  \\
  F_{01} & = & \frac{Q_e}{r^2} e^{2\phi} ,  \\
  F_{23} & = & \frac{Q_m}{r^2} . 
\end{eqnarray}
The properties of the CLH solutions are characterized by the mass 
$ M $, electric charge $ Q_e $, magnetic charge $ Q_m $ and asympotic value  
of the dilaton $ \phi_0 $. These solutions are related to the GHS solutions by a
coordinate transformation 
\begin{equation}
  \stackrel{\sim}{r} = \sqrt{r^2+\frac{Q_e^4}{4M^2}e^{4\phi_0}} + \frac{Q_e^2}{2M}e^{2\phi_0} ,
\end{equation}
as the magnetic charge $ Q_m $ is set to be zero. 
This new coordinate remove the peculiar singularity $ \stackrel{\sim}{r} = \frac{Q_e^2}{M} e^{2\phi_0} $ 
whose area is zero, to the essential singularity $ r = 0 $.

    Recently, the energy distribution according to the 
Einstein's energy - momentum pseudotensor was studied. Based on the GHS 
solutions, Virbhadra et. al.~\cite{6} found a charge independent result 
\begin{equation}
  E(r) = M,
\end{equation}
in which the positive energy is confined to the interior of the black
hole. On the other hand, based on the CLH solutions, we obtained a charge dependent
result~\cite{7} which is different from Virbhadra's 
\begin{equation}
  E(r) = M + \frac{M\lambda^2}{r^2} - \frac{1}{2\sqrt{r^2+\lambda^2}} \left( \frac{\beta\lambda^2}{r^2} + \lambda^2 + \beta \right) , \\
\end{equation}
By comparing Eq.(16) and (17), we find that the different 
coordinates choosen will induce the same total energy
but differents energy distributions without any relation.
This shortcoming can be overcome by using M{\o}ller's energy - momentum
pseudotensor.
   
    M{\o}ller's energy - momentum pseudotensor~\cite{2} is
\begin{equation}
  \Theta^\mu_\nu = \frac{1}{8\pi}\frac{\partial \chi^{\mu\sigma}_\nu}{\partial x^\sigma},
\end{equation}
where
\begin{equation}
  \chi^{\mu\sigma}_{\nu} = \sqrt{-g}(\frac{\partial g_{\nu\alpha}}{\partial x^\beta} - \frac{\partial g_{\nu\beta}}{\partial x^\alpha})g^{\mu\beta}g^{\sigma\alpha}.
\end{equation} 
The Greek indices run from 0 to 3 and $ x^0 $ is time coordinate. Then,
the energy component $ E $ is given by 
\begin{eqnarray}
  E & = & \int \int \int \Theta_{0}^{0} dx^{1} dx^{2} dx^{3} \nonumber \\
    & = & \frac{1}{8\pi} \int \int \int \frac{\partial \chi^{0k}_0}{\partial x^k} dx^1 dx^2 dx^3,  
\end{eqnarray}
where the Latin index takes values from 1 to 3. The M{\o}ller's
energy - momentum pseudotensor which differ from Einstein's 
energy - momentum pseudotensor is not necessary to carry out the
calculation in the quasi - Cartestian coordinate, so we can calculate
in the spherical coordinate.

    In the case of the GHS solutions, we obtain the nonvanishing components $ \chi^{0k}_0 $
in Eq.(20),
\begin{equation}
  \chi^{01}_0 = (2M - \frac{2M\alpha}{\stackrel{\sim}{r}})sin\theta . \\
\end{equation}
Applying the Gauss theorem, and plugging (21) into (20), we 
evaluate the integral over the surface of a sphere with radius $ r $.
\begin{equation}
  E(r) = \frac{1}{8\pi} \oint_{r} (2M - \frac{2M\alpha}{r})sin\theta d\theta d\varphi.
\end{equation}
Finally, for the GHS solution, we find that the energy within a sphere with radius $ r $ is
\begin{equation}
  E(r) = M - \frac{Q_e^2}{\stackrel{\sim}{r}} e^{2\phi_0}.  
\end{equation}

    In the case of the CLH solutions, 
the nonvanishing components $ \chi^{0k}_0 $ in (19) are 
\begin{equation}
  \chi^{01}_0 = (2M + \frac{4M\lambda^2}{r^2} - \frac{2\beta}{r^2}\sqrt{r^2 + \lambda^2}) sin\theta .\\
\end{equation}
For the CLH solution, the energy within a sphere with radius $ r $ is
\begin{equation}
  E(r) = M + \frac{2M\lambda^2}{r^2} - \frac{\beta}{r^2}\sqrt{r^2 + \lambda^2}.
\end{equation}
The energy distributions are shared both by the interior and by the exterior of  
those charged dilaton black hole.
We plot the energy distributions of the dyonic black holes or the
extremal dyonic black holes by "GNUPLOT". For the dyonic black hole or the extremal dyonic
black hole, see Fig.1 and Fig.4, we find that the energy distribution
can be positive or negative, but they are both positive in the 
region $ r > r_H $. For the pure electric or pure magnetic charged 
black hole, i.e. $ Q_m = 0 $ or $ Q_e = 0 $, we find the remarkable
property that the energy distributions are always positive except at 
singular point $ r = 0 $, see Fig.2,3,5,6. These results indicate that 
the physical charged black hole solution is either pure electric or pure
magnetic when the positive definite condition, all the energy distribution
function are positive definite expect the singularity, is imposed.

    Comparing the M{\o}ller' energy distributions of the GHS solutions 
and the CLH solutions, we find that the results of the CHL solutions
seem to be different from the GHS solutions. But they are related by
scratching the magnetic charge $ Q_m $ of CHL solution and by the coordinate
transformation (15) which is a purely spatial transformation. Therefore,
it will be the same energy distributions of the CLH
solutions and the GHS solutions. Then we confirm that the statement in
M{\o}ller's paper,
"the property of the M{\o}ller's energy complex is that 
transforms as 3 - scalar with respect to the 
group of purely spatial transformation", is still valid for the dilaton
gravity theory.
   
\begin{center}
{\bf Acknowledgements}
\end{center}
I.C. Yang would like to thanks Prof. K.S. Virbhadra and
Prof. J.M. Nester for useful comments and discussions.
This work is supported in part by the National Science Council of the 
Republic of China under grants NSC-86-2112-M006-003.

\newpage
\begin{figure}[hp]
   \input{fig1.tex}
   \caption{ The energy distribution of dyonic black hole with $ \phi_0 = 0 $, $ M = 2 $, $ Q_e = 1 $ and $ Q_m = 1 $.}
   \vspace{5mm}
   \input{fig2.tex}
   \caption{ The energy distribution of pure electric black hole with $ \phi_0 = 0 $, $ M = 2 $, $ Q_e = 1 $ and $ Q_m = 0 $.}
\end{figure}
\newpage
\begin{figure}[hp]
   \input{fig3.tex}
   \caption{ The energy distribution of pure magnetic black hole with $ \phi_0 = 0 $, $ M = 2 $, $ Q_e = 0 $ and $ Q_m = 1 $.}
   \vspace{5mm}
   \input{fig4.tex}
   \caption{ The energy distribution of extremal dyonic black hole with $ \phi_0 = 0 $, $ Q_e = \sqrt{2} $ and $ Q_{m} = \sqrt{2} $.}
\end{figure}
\newpage
\begin{figure}[hp]
   \input{fig5.tex}
   \caption{ The energy distribution of extremal electrically charged black hole with $ \phi_0 = 0 $, $ M = 2 $, $ Q_e = 0 $ and $ Q_m = 2\sqrt{2} $.}
   \vspace{5mm}
   \input{fig6.tex}
   \caption{ The energy distribution of extremal magnetically charged black hole with $ \phi_0 = 0 $, $ M = 2 $, $ Q_e = 2\sqrt{2} $ and $ Q_m = 0 $.}  
\end{figure}

\end{document}